\documentclass[preprint,showpacs,aps]{revtex4}
\usepackage{graphicx}
\begin{document}

\title{Multicomponent dense electron gas as a model of Si MOSFET}
\author{S. V. Iordanski}
\author{A. Kashuba}

\affiliation{L. D. Landau Institute for Theoretical Physics, Russian Academy of Sciences, 2 Kosygina st., 119334 Moscow}

\date{\today}

\begin{abstract}
We solve a 2D model of $N$-component dense electron gas in the limit $N\to \infty$ and in a range of the Coulomb interaction parameter: $N^{-3/2}\ll r_s\ll 1$. The quasiparticle interaction on the Fermi circle vanishes as $\hbar^2/Nm$. The ground state energy and the effective mass are found as series in powers of $r_s^{2/3}$.  In the quantum Hall state on the lowest Landau level at integer filling: $1\ll\nu<N$, the charge activation energy gap and the exchange constant are: $\Delta=\log(r_s N^{3/2})\hbar\omega_H/\nu$ and $J=0.66 \hbar\omega_H/\nu$.
\end{abstract}


\maketitle

Two-dimensional electron gas (2DEG) in GaAs quantum wells and Si heterostructures \cite{ando} is a unique system where density of electrons $n$ can be varied widely. Effects of electrostatic Coulomb interaction are determined by the variable dimensionless coupling: $r_s=e^2m/\sqrt{\pi n}\hbar^2$. In Si MOSFET interesting phenomena was observed recently at relatively large $r_s^\dagger\sim 10$: a conductivity 'fan' similar to the metal-insulator transition \cite{ak}, a sharp increase of the effective electron mass and magnetic susceptibility \cite{pud}. At large $r_s$ no exact model exists and many phenomenological approximations have been developed.

At small $r_s$ there is an exact model - dense electron gas \cite{GMB} but two of its key predictions have not been confirmed in the experiment. First, theory predicts a reduction of the effective mass: $m^*/m-1\sim r_s\log(1/r_s)$ at $r_s\to 0$ \cite{gm} - because the forward scattering is larger then the backward scattering for Coulomb potential. Second, in the integer quantum Hall ferromagnet state on the lowest Landau level, theory \cite{bie} predicts the Coulomb charge activation gap $\Delta\sim e^2\sqrt{eH/\hbar c}$ in strong magnetic field $H$ ($r_s\to 0$). Yet, in Si(100) MOSFET the Shubnikov-de Haas oscillation experiments \cite{ss,pud} find: $m^*/m= 1+0.08 r_s$, at $3>r_s>0.9$, and the magnetocapacitance experiment \cite{dol} finds much lower charge activation gaps: $\Delta\sim \hbar eH/mc$, at $4>r_s>1.5$. In phenomenological models (see Refs. in \cite{ando}) experimental $m^*(r_s)>m$ was reproduced in the large $r_s$ domain with a crossover at $r_s^*\sim 1$ from the theory \cite{GMB} with $m^*<m$ to the large $r_s$ with $m^*>m$. Experiments \cite{ss,pud} bound this crossover to be small $r^*_s<0.9$. But perturbation series in one parameter $r_s$ could hardly have the two so widely different crossovers: $r_s^\dagger$ and $r_s^*$.

In this Letter we show that a systematic model of multicomponent dense electron gas agrees with the experiment. In Si electron states have a valley degeneracy \cite{ando}, corresponding to different band minima. For (100) orientation of 2D plane there is $N=4$ equivalent spin-valley states. Two valley band states here differ only by plane-waves $\exp(\pm iQz)$ in the perpendicular direction with an atomic value of momentum $Q$. These states are strongly orthogonal. For (111) orientation the spin-valley degeneracy is especially large $N=12$, but in experiments only $N=4$ states are filled with electrons \cite{ando} - a long standing puzzle. Electron gas in the limit $N\to\infty$ was first considered in Ref.\cite{t} and the leading term of the ground state energy was shown to coincide with that of a dense charged boson gas \cite{fb}. 

For $N$-component dense 2DEG in the limit $N\to\infty$ we find a crossover at $r_s^*=1/N^{3/2}$ from the theory \cite{GMB} at vanishing $r_s$ to a novel systematic theory at:
\begin{equation}\label{cond1} r_s\ll 1 \quad {\rm and} \quad \label{cond2} r_s\gg r_s^*=1/N^{3/2}.
\end{equation}
It describes a system of interacting electrons and plasmons. Plasmons are excitations with large characteristic momentum $q_0\gg p_F$ and energy $\omega_0 \gg \epsilon_F$. Landau theory of Fermi liquid uses only one function - the amplitude of quasiparticles scattering - to derive all Landau Fermi liquid parameters. In our theory the interaction between quasiparticles vanishes as $1/Nm$ and the electron subsystem is the ideal Fermi gas. The exchange of high energy plasmons gives rise to a polaronic like renormalization of the effective electron mass as a series in powers of $r_s^{2/3}$. In the case of integer quantum Hall state on the lowest Landau level we predict a linear dependence on magnetic field of the charge activation energy gap and the exchange constant that agree better with the experiments.

2DEG Hamiltonian is expressed in terms of the second quantized electron operators \cite{agd}:
\begin{eqnarray}\label{FGH} 
\hat{H}=\frac{1}{2m}\int \psi^\dagger_\alpha(\vec{r})  \left(-i\hbar\vec{\nabla}+\frac{e}{c}\vec{A}(\vec{r})\right)^2 \psi_\alpha(\vec{r}) \ d^2\vec{r}+ \nonumber\\  +\frac{1}{2}\int\!\!\int \frac{e^2}{|\vec{x}-\vec{r}|} \psi^\dagger_\alpha(\vec{x}) \psi^\dagger_\beta(\vec{r})  \psi_\beta(\vec{r}) \psi_\alpha(\vec{x})\ d^2\vec{x}\  d^2\vec{r},
\end{eqnarray}
where $\alpha,\beta=1..N$ are spin-valley indices. $N$ is an even length of fermion spinor. No valley mixing in the density operators, isotropic mass tensor and a positive uniform charge at large distance $d$ from the 2D plane are assumed. At first we assume zero magnetic field and $\alpha_F=e^2/\hbar v_F\approx r_s\sqrt{N}\ll 1$ instead of the first condition (\ref{cond1}) and then we extend the theory to all $r_s\ll 1$. We use Matsubara diagrammatic expansion in terms of the Coulomb interaction. New Coulomb line multiplies the diagram by a small parameter $\alpha_F$, whereas each electron loops brings a large factor $N$ with it. Therefore, the leading order is given by diagrams with minimum number of Coulomb lines per electron loop, with an essential block being the RPA screened Coulomb interaction - a sequence of alternating Coulomb lines and electron loops \cite{pn}. The second condition (\ref{cond1}) makes a typical RPA momentum to be large: $q\sim q_0\gg p_F$, where $q_0=(8\pi e^2 mn)^{1/3}$ $= r_s^{1/3}\sqrt{N} p_F$ and $\omega_0=q_0^2/\sqrt{2}m$ defines a typical plasmon momentum and energy. The electron polarization operator is:
\begin{equation}\label{PolOp} 
\Pi(\omega,\vec{q})=\frac{2n \epsilon(\vec{q})} {\omega^2+ \epsilon^2(\vec{q})},
\end{equation}
where $\epsilon(\vec{q})=q^2/2m$ is the electron dispersion. It depends on the total density $n$ and is independent of an electron distribution over $N$ spin-valleys as long as $p_{F\alpha}\ll q_0$. The RPA Coulomb interaction is:
\begin{equation}\label{Plas} 
D(\omega,\vec{q})=\frac{2\pi e^2}{q} \left(1+\frac{2\pi e^2}{q}\Pi(\omega,\vec{q})\right)^{-1} .
\end{equation}
Its pole corresponds to a plasmon excitation with a dispersion: $\omega(\vec{q})= \sqrt{q^4+q^3_0q}/2m$. Electrons can be integrated out from the model (\ref{FGH}) leaving an effective theory of plasmons with typical momentum $q_0$ and a propagator (\ref{Plas}) which weakly interact (in the limit $r_s\ll 1$) due to three plasmon, four plasmon etc vertices. A plasmon vertex with $k$ leg with external momenta $q_i\sim q$ and frequencies $|\omega_i|\sim \omega$ vanishes in the limit $q_i^2/m\ll |\omega_i|$ as $nq^2/m\omega^k$ for $k$ even and $nq^4/m^2 \omega^{k+1}$ for $k$ odd due to the electron number conservation.

\begin{figure}
\includegraphics{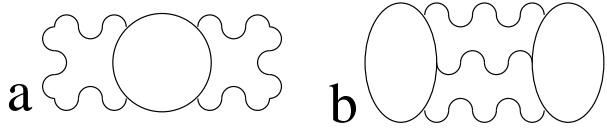}
\caption{\label{fig} a) Four-leg plasmon vertex and b) two three-leg plasmon vertices contribution to the energy. Solid lines are electron propagations and wavy lines are plasmon propagations.}
\end{figure}

The kinetic energy density $E_{kin}=\pi n^2/Nm$ is small because electrons are evenly distributed over $N$ spin-valley states. In the leading order the energy of 2DEG is given by the zero energy of plasmons: $E_0=E_{HF}+E'$, where $E_{HF}=-8r_sn^2/3m\sqrt{N}$ is the 2D Hartree-Fock exchange energy and $E'$ is the RPA energy \cite{pn}:
\begin{equation}\label{E0} 
E'\!=\!\!\int\left[\log\left(1+\frac{2\pi e^2}{q}\Pi(\omega,\vec{q})\right) -\frac{2\pi e^2}{q} \Pi(\omega,\vec{q})\right] \frac{d\omega\,d^2\vec{q}}{2(2\pi)^3}.
\end{equation}
At $q\sim p_F$ we single out from $E'$ a term that cancels $E_{HF}$ exactly and the remaining plasmon energy (also called a correlation energy) can be evaluated as \cite{com}:
\begin{eqnarray} \label{Ecor} 
E_0= -\ \frac{3\sqrt{\pi}}{4} \Gamma\left(\frac 23\right) \Gamma\left(\frac 56\right) r_s^{4/3}\frac{n^2}{m}. 
\end{eqnarray}
Plasmon energy (\ref{Ecor}) coincides in the leading order with the ground state energy of a dense charged boson gas \cite{fb}. In the case of anisotropic electron mass tensor of Si(111) we compare two states with i) all $N=12$ states being filled and ii) only $N=4$ states related by the spin and time reversal being filled. Because $\log$ in (\ref{E0}) is a convex function the plasmon energy (\ref{Ecor}) in the case ii) is always lower then in the case i) by an amount $\approx 0.04 r_s^{4/3}n^2/\langle m\rangle$. This can overcome the kinetic energy difference: $\pi n^2/6\langle m\rangle $ at experimental densities $r_s>1$.

The next corrections to the plasmon energy are given by the second order diagram a) and the third order diagram b) on Fig.\ref{fig}. Provided the internal frequency and momentum are related by plasmon dispersion, the value of a diagram is proportional to $r_s^{2/3}$ in power of $N_i-N_L+2$, where $N_i$ is the number of Coulomb lines and $N_L$ is the number of electron loops. We have evaluated numerically two diagrams on Fig.\ref{fig} and find: $E_0+E_1=-(2.03191\ r_s^{4/3}-0.156(1)r_s^2)\ n^2/m$, where the first term represents Eq.(\ref{Ecor}). Comparing vertices of the fermion (\ref{FGH}) and the dense charged bose gas models we conclude that these models are different.

\begin{figure}
\includegraphics{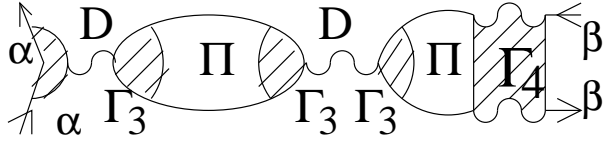}
\caption{\label{fig2} A particle-hole ladder for vertex $\Gamma$. Two leg element $D$ three leg element $\Gamma_3$ four leg elements $\Gamma_4$ is shown.}
\end{figure}

\begin{figure}
\includegraphics{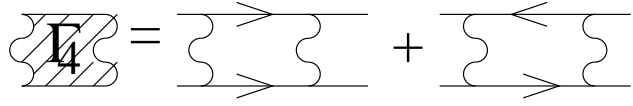}
\caption{\label{fig1} Four-leg vertex is a sum of two diagrams in the leading order.}
\end{figure}

A specific feature of our mean field theory that distinguishes it from the standard Landau Fermi liquid theory is that electron and plasmon excitations are found in a wide momenta range $p_F\ll p\ll q_0$. Let us prove that the four fermion vertex \cite{agd} vanishes on the infrared side of this range: $\Gamma_{\alpha\alpha; \beta\beta} (\epsilon_i,p_i) \sim Q^2/mn$ at $p_i/q_0\to 0$, where $\epsilon_i\sim p_i^2/2m$ and $i=1..4$. $\Omega=\epsilon_1- \epsilon_2$ and $\vec{Q}= \vec{p}_1-\vec{p}_2$ are the transfered frequency and momentum in the particle-hole channel on Fig.\ref{fig2}. A pair of electron and hole Green functions with independent integration over frequency and momentum and sum over spin-valley index $\gamma$ gives a large term \cite{agd}. Therefore vertex $\Gamma(\Omega,Q)$ is a ladder of alternating pairs - polarization operators $\Pi(\Omega,Q)$ - and blocks $M$, defined as a set of all diagrams that can not be cut over the pair lines. The block $M$ is assumed to be independent of the internal integration momenta and frequency of the adjacent $\Pi(\Omega,Q)$. The block $M$ can be further divided into i) a part that can be cut over one plasmon line: $\Gamma_3 D(\Omega,Q) \Gamma_3$, where $\Gamma_3$ is a three leg one-plasmon irreducible part at vanishing leg momenta, and ii) a four leg one-plasmon irreducible part $\Gamma_4$ at vanishing leg momenta. Let $\Gamma_D(\Omega,Q)$ and $\Gamma_\Gamma(\Omega,Q)$ be the parts of the total vertex $\Gamma(\Omega,Q)$ that end on the left with a plasmon line $D$ or with a block $\Gamma_4$. The Dyson equation for the particle-hole channel is algebraic:
\begin{equation}\label{DysonG}
\left(\begin{array}{c} \Gamma_D \\ \Gamma_\Gamma \end{array} \right)=\left(\begin{array}{c} D\Gamma_3 \\ \Gamma_4 \end{array} \right) -\left(\begin{array}{cc} \Pi D (\Gamma_3^2-1) & D\Gamma_3\Pi \\ \Gamma_4\Pi\Gamma_3 & \Gamma_4\Pi \end{array} \right) \left(\begin{array}{c} \Gamma_D \\ \Gamma_\Gamma \end{array} \right).
\end{equation}
The total four fermion vertex is given as $\Gamma=\Gamma_3\Gamma_D+\Gamma_\Gamma$:
\begin{equation}\label{Gamm} 
\Gamma=\frac{D\Gamma_3^2+\Gamma_4(1-D\Pi)}{\Gamma_3^2+(1+\Gamma_4\Pi)(1-D\Pi)}.
\end{equation}
If $\Omega\sim Q^2/m$ then we estimate the plasmon propagator $D(\Omega,Q) \approx \Pi^{-1}(\Omega,Q)\sim Q^2$ to be small but the factor: $(1-D\Pi)\sim Q^3$, is even smaller in the limit $Q\to 0$. We use the Ward identity: $\Gamma_3=Z^{-1}$, where $Z<1$ is the Green function pole renormalization. $\Gamma_4$ is given in the lowest order by the two diagrams on Fig.\ref{fig1}:
$\Gamma^0_4= -\sqrt{\pi}\Gamma(2/3)\Gamma(5/6)r_s^{4/3}/\sqrt{3}m$. It represents a retarded attractive interaction because it appears in the second order of perturbation in plasmon exchange. Assuming that $\Gamma_4$ is finite we conclude from (\ref{Gamm}) that $\Gamma(0,Q)=D(0,Q) =Q^2/4m^*n Z^2$. 

This suggests an analogy between Fermi gas and a ferromagnet with spontaneously broken symmetry where both spin-wave dispersion and interaction vanish as $Q^2$ according to the Goldstone theorem. A continuation to momenta on the Fermi circle gives the quasiparticle interaction: $\Gamma=2\pi/mN$.

There is no infrared divergences in the block $\Gamma_4$ despite of the possibility that some internal frequency integration are determined by electron energy poles rather then by plasmon energy poles in the momentum range: $p_F\ll p\ll q_0$. Would be divergent are those diagrams that i) can be cut into two or more parts over only electron lines; ii) each part, emerged as a result, is one-plasmon irreducible after all internal electron lines are contracted into points. These parts where all plasmon lines are closed into loops are fermion vertices symbolically written as $V_{2k}\int \psi^{\dagger k}\psi^k dx^2dt$. In the Cooper channel the four-leg vertex $V_4=\Gamma\sim Q^2$ vanishes thus cutting off a specific 2D logarithmic divergence \cite{agd}. All higher order vertices are regular as functions of leg momenta and are irrelevant in 2D as a simple power counting shows: $t\sim x^2$ and $\psi\sim 1/x$.

Despite the vanishing interaction between electrons there is an essential plasmon polaronic effect. The electron self-energy \cite{agd}:
\begin{equation}\label{Dyson}
\Sigma(\epsilon,\vec{p})= - \int D(\omega,\vec{q}) G(\epsilon+\omega,\vec{p}+\vec{q})\  \frac{d\omega\ d^2\vec{q}}{(2\pi)^3},
\end{equation}
is related to the electron Green function: $G^{-1}(\epsilon,\vec{p})=i\epsilon - p^2/2m+\mu- \Sigma(\epsilon,\vec{p})$, where $\mu$ is the chemical potential. At $\epsilon\ll \omega_0$ and $p\ll q_0$, we calculate the self-energy (\ref{Dyson}) and find: $G^{-1} (\epsilon,\vec{p})= Z(i\epsilon- \epsilon(\vec{p}))^{-1}$, where the quasiparticle dispersion is: $\epsilon(\vec{p})=(p^2-p_F^2)/2m^*$, with the renormalized effective mass:
\begin{equation}\label{MassRen} \frac{m}{m^*}=1- \frac{1}{10\sqrt{\pi}}\ \Gamma\left(\frac 13\right)\Gamma\left(\frac 76\right)\ r_s^{2/3} ,
\end{equation} 
and $Z^{-1}=1+\Gamma(1/3)\Gamma(7/6)\ r_s^{2/3}/2\sqrt{\pi}$ is a renormalization of the Green function pole. 

A static screened potential of an external charge $z$ immersed into 2DEG is given by $D(0,\vec{q})$ (\ref{Plas}): $V(\vec{q})= 2\pi e^2 zq^2/ (q^3+q_0^3)$. It grows with the transfered momentum that helps to explain the heavier effective mass (\ref{MassRen}) because the backward scattering is larger than the forward scattering. 

The temperature dependence of the effective mass comes from the momentum range: $p_F\ll p\ll q_0$. Taking into account the temperature dependence of the polarization operator in the equation for the self-energy (\ref{Dyson}) we encounter logarithmic infrared divergence. Deriving and solving a simple renormalization group equation we find a growing effective mass if momentum of the quasiparticle decreases. On the Fermi line we find:
\begin{equation}\label{Temp} 
\frac{1}{m_*^2(T)}=\frac{1}{m_*^2}-\frac{NT^2}{12 n^2}\ \log \frac{q_0}{p_F},
\end{equation}
where zero temperature effective mass $m_*$ is given by (\ref{MassRen}). In the experimental situation of Si(100) MOSFET the effective mass (\ref{Temp}) could strongly depend on temperature.

At $T\ll\omega_0$ plasmon thermal fluctuations freeze out whereas the ideal electron subsystem evolves from a degenerate gas at $T\ll\epsilon_F$ to the Boltzmann gas at $T\gg\epsilon_F$. 2DEG thermodynamic potential: $\Omega(T)=N \Omega_0(T,\mu)+ E_0+E_1$, where $\Omega_0(T,\mu)$ is the thermodynamic potential of the ideal Fermi gas with a property: $N\Omega_0(0,\epsilon_F)= E_{kin}$.

\begin{figure}
\includegraphics{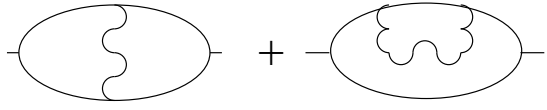}
\caption{\label{fig3} Two diagrams for a spin susceptibility correction}
\end{figure}

Static magnetic susceptibility is related to a change of the total energy in external magnetic field: $\delta E=-N\chi^* (\mu_B H)^2/2$. $\chi^*$ is given by the Pauli ideal gas susceptibility: $\chi=m/2\pi$, and by two diagrams on Fig.\ref{fig3}. Evaluating them we find: $\chi^*/\chi=m^*/m$. There is no exchange part coming from the interaction between quasiparticles. Thus, there is a single Landau Fermi liquid parameter $m^*/m$ given by a series in powers of $r_s^{2/3}$. We observe that the experimental effective mass \cite{ss,pud} agrees well with (\ref{MassRen}). The experimental susceptibility \cite{oka,pud} is larger then $\chi^*$ that indicates some exchange effects for $N=4$.

In the end we consider the quantum Hall state \cite{PG}, with an integer filling factor $\nu$, defined as a ratio of electron number to the magnetic flux number. If $1\ll\nu< N$ then $\nu$ is a number of occupied spin-valley states on the lowest Landau level. This state is degenerate under global rotations in electron Fock space by unitary matrices from the Grassmanian coset: $F=U(N)/ U(\nu) \otimes U(N-\nu)$, that corresponds to a ferromagnetic order parameter at zero temperature. We use the so-called magnetic units: $\hbar=1$, $e=c$, magnetic length $l_H=1$, cyclotron frequency $\omega_H=eH/mc=1/m$ and $r_s=\sqrt{2}e^2/\omega_H l_H\sqrt{\nu}$. The polarization operator is:
\begin{equation}\label{pol} 
\Pi(\omega,\vec{q})=\frac{\nu}{2\pi}\sum_{s=1}^{\infty}\frac{q^{2s}}{2^s\ s!} \frac{2s\omega_H} {\omega^2+\omega_H^2 s^2} \exp\left(-\frac{q^2}{2}\right),
\end{equation}
and at $ql_H\gg 1$ it transforms into (\ref{PolOp}). We find the self-energy ({\ref{Dyson}) using the Green function: $G^{-1}(\epsilon)= i\epsilon+\mu$, on the lowest Landau level. The difference between the self-energies for electron in the empty spin-valley state and electron in the occupied spin-valley state gives the charge activation gap: $\Delta=\Sigma_e-\Sigma_o$. We single out the Hartree-Fock term: $E_{HF}=\sqrt{\pi/2}\, e^2/l_H$, and find:
\begin{equation} \label{S} 
\Delta=E_{HF}-\int\frac{d\omega\, d^2\vec{q}} {(2\pi)^3} \left(\frac{2\pi e^2}{q}- D(\omega,q)\right) \frac{2\mu}{\omega^2+\mu^2} ,
\end{equation}
where only the real part of electron Green function is essential. $\mu=\Delta/2$ is the chemical potential inside the ferromagnetic gap and the last fraction in (\ref{S}) becomes approximately delta-function $\delta(\omega)$, if $\mu\ll\omega_H$. Hartree-Fock term cancels out exactly like in zero magnetic field and after frequency integration we find:
\begin{equation} \label{gapf} 
\Delta=\frac{\omega_H}{2}\int_0^\infty D(q)dq= \frac{\hbar\omega_H}{\nu}\left( \log(r_s \nu^{3/2})+0.277\right),
\end{equation}
where $D^{-1}(q)=e^{q^2/2}/r_s\sqrt{2\nu}+\nu\sum_{s=1}^\infty q^{2s-1}/2^ss!s$. Note that $\Delta\ll E_{HF}$. The gap on the lowest Landau level (\ref{gapf}) is similar to the gap in the $N=2$ quasi-classical case of weak magnetic field and odd integer $\nu\gg 1$ \cite{ag}.

We find a dispersion of spin waves as a pole of correlation function $C(\vec{x}-\vec{r})=\langle\psi^\dagger_e(\vec{x}) \psi_o(\vec{x}) \psi^\dagger_o(\vec{r}) \psi_e(\vec{r})\rangle$, which is uniform because it describes a neutral excitation. It is given by a ladder set of diagrams like the second one on Fig.\ref{fig1}, where electron Green function include the self-energy: $G^{-1}_{eo}=i\epsilon \pm \Delta/2$. Using Landau gauge representation of the density operators \cite{bie} we find:
\begin{equation}\label{ladder} 
C(\Omega,Q)=\sum_{k=0}^\infty \left(-\frac{\omega_H}{i\Omega-\Delta}\int D(q)\exp(i\vec{Q}\times\vec{q})\frac{d^2\vec{q}}{4\pi q}\right)^k
\end{equation}
a sum over $k$-leg ladder diagrams. It has a pole at $i\Omega=\epsilon(Q)$ with the spin-wave dispersion:
\begin{equation}\label{SW} \epsilon(\vec{Q})=\frac{\omega_H}{2}\int_0^\infty
D(q) \left(1-J_0(qQ)\right)\ dq.
\end{equation}
At intermediate wavelengths $r_s\nu^{3/2}\ll Ql_H\ll 1$ spin wave dispersion is logarithmic: $\epsilon(Q)=-\omega_H\log(Ql_H)/\nu$, whereas in the long wavelength limit at $Ql_H\ll r_s\nu^{3/2}$ we recover Goldstone dispersion: $\epsilon(Q)=JQ^2$, where the exchange constant can be evaluated numerically:
\begin{equation} \label{ExC} J=\frac{\omega_H}{8}\int_0^\infty q^2 D(q) dq= 0.6613\,\frac{\omega_H}{\nu}
\end{equation} 
A pair of Skyrmion topological defects of ferromagnetic order \cite{skkr} has a lower activation energy: $\Delta=J$, then the electron-hole pair (\ref{gapf}).

We are greatfull to G. M. Eliashberg who shaped very much the content. This work was supported by Russian Foundation for Basic Research grant 01-02-17520a and INTAS grant 99-01146.


\begin{thebibliography}{99}

\bibitem{ando}  T. Ando, A. B. Fowler, and F. Stern, Rev. Mod. Phys. {\bf 54}, 437-672 (1982)
\bibitem{ak} E. Abrahams, S. V. Kravchenko and M. P. Sarachik, Rev. Mod. Phys. {\bf 73}, 251 (2001)
\bibitem{pud} V. M. Pudalov, M. E. Gershenson, H. Kojima et al, Phys. Rev. Lett. {\bf 88}, 196404 (2002)
\bibitem{GMB} M. Gell-Mann and K. A. Brueckner, Phys. Rev. {\bf 106}, 364 (1957); K. Sawada, Phys. Rev. {\bf 106}, 372 (1957)
\bibitem{gm} M. Gell-Mann, Phys. Rev. {\bf 106}, 369 (1957)
\bibitem{bie} Yu. A. Bychkov, S. V. Iordanski and G. M. Eliashberg, JETP Lett.
{\bf 33} 152 (1981); C. Kallin and B.I. Halperin, Phys. Rev. B {\bf 30}, 5655 (1984)
\bibitem{ss} J. L. Smith and P. J. Stiles, Phys. Rev. Lett. {\bf 29}, 102 (1972)
\bibitem{dol} V. S. Khrapai, A. A. Shashkin and V. T. Dolgopolov, cond-mat/0202505
\bibitem{t} Y. Takada, Phys. Rev. B {\bf 43}, 5962 (1991)
\bibitem{fb} L. L. Foldy, Phys. Rev. {\bf 124}, 649 (1961)
\bibitem{agd}  A. A. Abrikosov, L. P. Gorkov and I. E. Dzyaloshinski,
{\it Methods of Quantum Field Theory in Statistical Physics}, (Dover New York, 1975)
\bibitem{pn} D. Pines and P. Nozieres, {\it The Theory of Quantum Liquids}, (Benjamin New York, 1966)
\bibitem{com} Capacitor energy density $e^2dn^2$ insures positive pressure and compressibility in the limit $d\sqrt{n}\to\infty$. 
\bibitem{oka} T. Okamoto, K. Hosoya, S. Kawaji and A. Yagi, Phys. Rev. Lett. {\bf 82}, 3875 (1999)
\bibitem{PG} {\it The Quantum Hall Effect}, edited by E. R. Prange and 
S. M. Girvin (Springer Verlag, New-York, 1990)
\bibitem{ag} I. L. Aleiner and L. I. Glazman, Phys. Rev. B {\bf 52}, 11296-11312 (1995)
\bibitem{skkr} S. L. Sondhi, A. Karlhede, S. A. Kivelson and 
E. H. Rezayi, Phys.Rev. B {\bf 47}, 16419 (1993)
\end{thebibliography}
\end{document}